\documentstyle[twoside,fleqn,espcrc2]{article}
\begin{document}

\title{A separate SU(2) for the third family: Topflavor}
\author{D. J. Muller and S. Nandi \thanks{%
Presented by S.~Nandi.}
\address{Department of Physics,
         Oklahoma State University,\\
         Stillwater, OK 74078}}

\begin{abstract}
We consider an extended gauge group for the electroweak interaction: SU(2)$%
_1 \times$ SU(2)$_2 \times$ U(1)$_Y$ where the first and second generations
of fermions couple to SU(2)$_1$ while the third generation couples to SU(2)$%
_2$. Bounds based on precision observables and heavy gauge boson searches
are placed on the new parameters of the theory, and the potential of the
theory to explain $R_b$ and $R_c$ is explored. We investigate processes that
could produce observable signals at the LHC and NLC.
\end{abstract}

\maketitle

\section{Introduction}

We investigate the possibility that the weak interactions of the third
generation of fermions are fundamentally different from the first and second
family. In particular, we look at the gauge group SU(2)$_1\times $ SU(2)$%
_2\times $ U(1)$_Y$ where the first and second generations couple to SU(2)$%
_1 $ and the third generation couples to SU(2)$_2$. We call this model
Topflavor in analogy to the Topcolor model \cite{topcolor}. Similar models
of generational nonuniversality have been considered in the past \cite
{similar}; here we consider the phenomenology of this gauge group and its
potential to help explain the $R_b$ and $R_c$ problem.

\section{The Model}

The left-handed first and second generation fermions form doublets under
SU(2)$_1$ and are singlets under SU(2)$_2$. Conversely, the left-handed
third generation fermions form doublets under SU(2)$_2$ and singlets under
SU(2)$_1$. All right handed fields are singlets under both SU(2)$_1$ and
SU(2)$_2$. With these representations the theory is anomaly free. The
covariant derivative is
\begin{equation}
D_\nu = \partial_\nu - i g_1 \stackrel{\rightarrow}{T} \cdot \stackrel{
\rightarrow}{W_\nu} - i g_2 \stackrel{\rightarrow}{T^{\prime}} \cdot
\stackrel{\rightarrow}{W^{\prime}_\nu} - i \frac {g^{\prime}}{2} Y B_\nu
\end{equation}
where the $W^a$ belong to SU(2)$_1$ and $W^{\prime a}$ belong to SU(2)$_2$.

The symmetry breaking in the theory is accomplished in two stages. In the
first stage the SU(2)$_1 \times$ SU(2)$_2$ is broken down to the SU(2)$_W$
of the standard model (SM). This is accomplished by introducing a Higgs
field $\Phi$ that transforms as a doublet under each SU(2) with the vev
\begin{equation}
\langle \Phi \rangle = \frac{1}{\sqrt{2}} \pmatrix{ u & 0 \cr 0 & u \cr } \
{}.
\end{equation}

In the second stage the remaining symmetry, SU(2)$_W \times$ U(1)$_{Y}$, is
broken down to U(1)$_{em}$. This is accomplished by introducing one or the
other of two Higgs fields that we call H$_1$ and H$_2$. Under (SU(2)$_1$,
SU(2)$_2$, U(1)$_Y$), H$_1$ transforms as (2, 1, 1) and obtains a vacuum
expectation value (vev) $\langle {\rm H}_1 \rangle = (0, v_1)$. H$_2$
transforms as (1, 2, 1) and develops a vev $\langle {\rm H}_2 \rangle = (0,
v_2)$.

The gauge bosons of the theory obtain mass through their interaction with
the Higgs fields. The fields in the current basis ($W_3$, $W^{\prime}_3$, $B$
) are related to the fields in the mass basis ($\gamma$, $Z_l$, $Z_h$)
through an orthogonal matrix, {\bf R}:
\begin{equation}
\pmatrix{ W_3 \cr W'_3 \cr B \cr } = {\rm {\bf R^\dagger}} (\phi, \theta_W,
\epsilon_1, \epsilon_2) \pmatrix{ \gamma \cr Z_l \cr Z_h \cr }
\end{equation}
where $Z_l$ is the $Z$ boson observed at present colliders and $Z_h$ is
called the ``heavy $Z$ boson''. Moreover, $\epsilon_1 = v_1^2/u^2$, $%
\epsilon_2 = v_2^2/u^2$, $\theta_W$ is the weak mixing angle, and $\phi$ is
an additional mixing angle such that the couplings of the theory are related
to the electric charge by $g_1 = \frac{e}{\cos \phi \sin \theta_W}$, $g_2 =
\frac{e}{\sin \phi \sin \theta_W}$, and $g^{\prime}= \frac{e}{\cos \theta_W}$
{}.

In the charged sector, the mass eigenstates are denoted by $W_l$ and $W_h$
where $W_l$ is the $W$ boson observed at present colliders and $W_h$ is
termed the ``heavy $W$ boson''. These are related to the current basis ($W$,
$W^{\prime}$) by an orthogonal matrix, {\bf R$^{\prime}$}:
\begin{equation}
\pmatrix{ W \cr W' \cr} = {\rm {\bf R^{\prime^\dagger}}} (\phi, \epsilon_1,
\epsilon_2) \pmatrix{ W_l \cr W_h \cr } \ .
\end{equation}

Due to the enlarged gauge and Higgs structure of this model, the couplings
of the particles are modified from their SM values. These modifications are
a function of $\phi $, $\epsilon _1$, and $\epsilon _2$, but in the limit of
$\epsilon _1=\epsilon _2=0$, the SM couplings of the fermions are recovered.
Since the couplings of the particles are modified from the SM values, the
phenomenology of this theory will be different from the SM. We investigate
the case where the couplings are perturbative. This places a restriction on
the allowed values that $\tan \phi $ can take: the requirements that $g_2^2/{%
4\pi }<1$ and $g_1^2/{4\pi }<1$ give $\tan \phi >0.2$ and $\tan \phi <5.5$,
respectively.

\section{Constraints From Experimental Data}

We restrict the possible values of the 3 new parameters ($\tan \phi$, $%
\epsilon_1$, and $\epsilon_2$) of the theory by requiring that the
theoretical predictions for various processes agree with experimental
values. Three versions of the model are considered:

\begin{description}
\item[case 1:]  Here we have just one Higgs doublet, $H_2$. In this case
there are only two parameters beyond the SM: $\tan \phi $ and $\epsilon _2$
or, more conveniently, $\tan \phi $ and $M_{W_h}$.

\item[case 2:]  Here we take only the other Higgs doublet, $H_1$. In this
case there are again only two parameters beyond the SM: $\tan \phi $ and $%
\epsilon _1$.

\item[case 3:]  Here we include both $H_1$ and $H_2$.
This is the general case where there are three parameters beyond the SM: $%
\tan \phi $, $\epsilon _1$, and $\epsilon _2$.
\end{description}

In case 1, we have that
\begin{equation}
\label{Rb1}\frac{\delta R_b}{R_b} = -0.8456 \frac{ \epsilon_2 \cos^4 \phi}
{(g_V^b)^2 + (g_A^b)^2}
\end{equation}
and
\begin{equation}
\label{Rc1}\frac{\delta R_c}{R_c} = 0.6912 \frac{ \epsilon_2 \cos^2 \phi
\sin^2 \phi} {(g_V^c)^2 + (g_A^c)^2}
\end{equation}
where $\delta R_b$ is the Topflavor model value of $R_b$ minus the SM value
and similarly for $R_c$. From the minus sign in eq. [\ref{Rb1}], we see that
the Topflavor model for this case gives a value for $R_b$ that is smaller
than the SM value which means that case 1 gives a value for $R_b$ that is
even farther from the experimental value than the SM. Similarly, we see from
eq. [\ref{Rc1}] that the case 1 value for $R_c$ is larger than the SM value
which, again, is in the opposite direction from the experimental value. Thus
case 1 does not provide an explanation for the $R_b$ or $R_c$ problems.

\begin{figure}[htb]
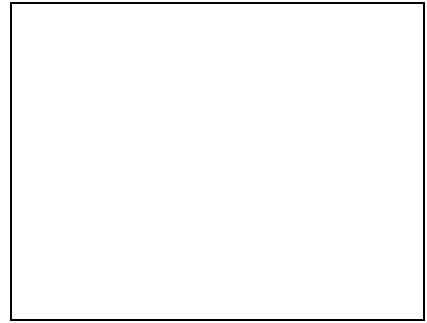

\framebox[55mm]{\rule[-21mm]{0mm}{40mm}}
\caption{Restriction curve for case 1.}
\label{restrict}
\end{figure}

We extensively considered the phenomenology of case 1 in another paper \cite
{topflavor}. In brief, the restriction imposed by various electroweak data
and heavy boson searches at hadronic colliders is presented in fig. \ref
{restrict}. The region above the line represents the allowed region in
parameter space while the region below the curve is disallowed. The left
hand side of the curve for $\tan \phi <2.1$ the largest restriction comes
from $\sigma _p^h$ and the experimental value for $\frac{g_\tau }{g_\mu }$
\cite{LEP}. The right side of the curve ($\tan \phi >2.1$) is obtained from
bounds set by search for a heavy W boson performed by the D0 collaboration
at Fermilab \cite{D0}.

For the second case, we have that
\begin{equation}
\label{Rb2}\frac{\delta R_b}{R_b} = 0.8456 \frac{\epsilon_1 \sin^2 \phi
\cos^2 \phi} { (g_V^b)^2 + (g_A^b)^2 }
\end{equation}
and
\begin{equation}
\label{Rc2}\frac{\delta R_c}{R_c} = -0.6912 \frac{\epsilon_1 \sin^4 \phi} {
(g_V^b)^2 + (g_A^b)^2 }
\end{equation}
as we see from eqs. [\ref{Rb2}] and [\ref{Rc2}], $R_b$ and $R_c$ for this
case move in the right direction although constraints from other observables
prevent this from being dramatic \cite{Rb}.

\begin{figure}[htb]
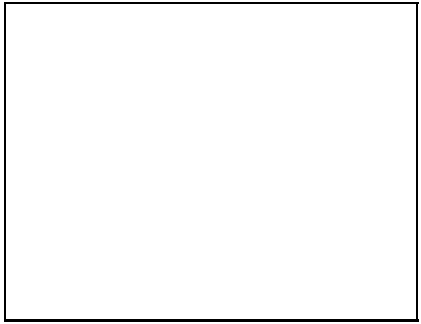

\framebox[55mm]{\rule[-21mm]{0mm}{40mm}}
\caption{Restriction curve for case 3.}
\label{restrict2}
\end{figure}

The third case, which is the general case with both Higgs fields present, is
considered in fig. \ref{restrict2} with the constraint $\frac{\epsilon_1}{
\epsilon_2} = 1.5$ (as an example). The solid line displays the 90\%
confidence level lower bound from a fit to various electroweak observables
(but not $R_b$ and $R_c$) and the $\tau$ lifetime \cite{LEP}. The effect of $%
\epsilon_1$ and $\epsilon_2$ on the couplings tends to cancel out at $\tan
\phi \simeq 0.8$ providing for the low dip in the allowed heavy $W$ mass.
The dotted line represents the restriction from the D0 search for a heavy $W$
boson \cite{D0}. $R_b$ can be accommodated to within $3\sigma$ here.

\section{Processes at Future Colliders}

First we consider NLC type lepton colliders. In the process $e^+ e^-
\rightarrow \mu^+ \mu^-$ there can be significant differences between the
Topflavor and SM predictions for the cross section due to the exchange of
the heavy $Z$ boson as well as to the deviation of the couplings from their
SM values. In fact, the coupling of the electron and muon to $Z_h$ goes as $%
\tan \phi$ to zeroth order in the $\epsilon$'s, so we expect the cross
section to increase significantly as $\phi$ increases. Consider the case 1
scenario (the results are very similar for all cases). For $\sqrt{s} = 1.8$
TeV, and a heavy Z mass of 850 GeV, the cross section increases from close
to the SM value at 3.8 pb to almost twice that value at 6.6 pb for $\tan
\phi = 5.5$. Thus the heavy Z boson will either be detected through this
process or the parameter space will be restricted further.

At the LHC there is the potential for single top production through the
production of a heavy $W$ boson and its subsequent decay to a top and bottom
quark. This decay process has a distinctive signature of 2 b-jets plus a
single large $p_T$ charged lepton. The decay cross section is about 1 nb for
a heavy $W$ mass of 1 TeV which is about the same as the cross section for
top pair production through normal SM processes.

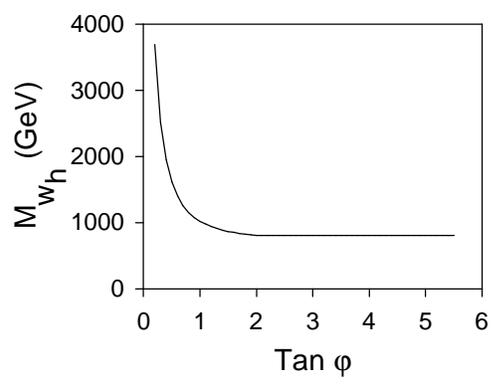
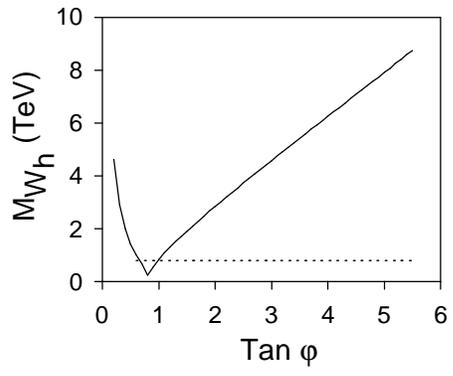
\end{document}